\documentclass{elsarticle}
\usepackage{setspace}

\usepackage[version=3]{mhchem} 
\usepackage{multirow}
\usepackage{comment}
\usepackage[utf8]{inputenc}
\usepackage{url}
\usepackage{setspace}
\usepackage{natbib}
\biboptions{compress}
\usepackage[dvipsnames]{xcolor}
\usepackage{soul}
\definecolor{myyellow}{RGB}{254, 252, 120}
\newcommand{\revision}[1]{#1}
\sethlcolor{myyellow}

\begin{document}

\begin{frontmatter}
\title{Transfer learning for solvation free energies: from quantum chemistry to experiments}
\author{Florence H. Vermeire}
\author{William H. Green \corref{cor1}}
\cortext[cor1]{Corresponding author\\Email address: \url{whgreen@mit.edu}}
\address[Massachusetts Institute of Technology]
{Department of Chemical Engineering, Massachusetts Institute of Technology, Cambridge, MA, 02139, U.S.A}

\begin{abstract}
Data scarcity, bias, and experimental noise are all frequently encountered problems in the application of deep learning to chemical and material science disciplines. Transfer learning has proven effective in compensating for the lack in data. The use of quantum calculations in machine learning enables the generation of a diverse dataset and ensures that learning is less affected by noise inherent to experimental databases. In this work, we propose a transfer learning approach for the prediction of solvation free energies that combines fundamentals from quantum calculations with the higher accuracy of experimental measurements \revision{using two new databases CombiSolv-QM and CombiSolv-Exp}. The employed model architecture is based on the directed-message passing neural network for the molecular embedding of solvent and solute molecules. 
A significant advantage of models pre-trained on quantum calculations is demonstrated for small experimental datasets and for out-of-sample predictions. The improved out-of-sample performance is shown for new solvents, for new solute elements, and for the extension to higher molar mass solutes. The overall performance of the pre-trained models is limited by the noise in the experimental test data, known as the aleatoric uncertainty. 
On a random test split, a mean absolute error of 0.21 kcal/mol is achieved. This is a significant improvement compared to the mean absolute error of the quantum calculations (0.40 kcal/mol). The error can be further reduced to 0.09 kcal/mol if the model performance is assessed on a more accurate subset of the experimental data. 
\end{abstract}


\end{frontmatter}

\section{Introduction}
Deep learning has emerged as an effective technique for property prediction in the field of chemical engineering and material science. In the last decade, many efforts have been made to replace structure-based estimation methods by deep neural networks \citep{Butler2018MachineScience, Toyao2020MachineProspects}. One major problem that is often encountered is data scarcity. Compared to other disciplines like image recognition and natural language processing, the availability and size of datasets in chemical engineering and material science are very limited.  
Transfer learning has been proposed as a technique to solve the problem of the low data regime \citep{Pan2010ALearning}. Success has been demonstrated in other disciplines, such as the transfer of knowledge from general image recognition to more specific medical imaging. 
Data scarcity is not the only problem related to the experimental nature of databases in chemical engineering and material science. They are often biased towards certain groups of components, cover only a limited domain of chemical space, and have an uncertainty associated with the experimental nature of the data. With transfer learning and the use of quantum chemical calculations, one can compensate for this bias and cover a larger chemical space by generating additional and diverse data.

The advantage of transfer learning with respect to data scarcity in chemical engineering and material science has been demonstrated in recent work. 
Within quantum machine learning, transfer learning has been used to calculate thermodynamic properties of molecules in vacuum at the coupled cluster level of theory, the gold standard of quantum chemical calculations. This has been done by Grambow et al. \cite{Grambow2019AccurateApproach} and Smith et al. \citep{Smith2019ApproachingLearning} Large DFT-based datasets have been used to pre-train models that were further fine-tuned on computationally expensive coupled cluster calculations. 
Ma et al. \citep{Ma2020TransferFrameworks} demonstrated the advantage of transfer learning for gas adsorption on metal organic frameworks. Parameters were transferred from a model trained on a large dataset of hydrogen gas adsorption at 100 bar and 243 K to initialize the parameters of a model fine-tuned at 130 K and for methane adsorption with a smaller dataset.
This technique has also been effective at transfering knowledge between disciplines of materials. Yamada et al. \citep{Yamada2019PredictingLearning} proposed a shot-gun transfer learning approach where models trained on small molecules were used to aid learning of polymer properties, and models trained on organic materials were used to aid learning of inorganic material properties.
Jha et al. \citep{Jha2019EnhancingLearning} used transfer learning to predict the formation enthalpy of crystal structures starting from the elemental composition. In their approach, a model was pre-trained on a large dataset of DFT calculations. All model parameters were used to initialize a new model that was fine-tuned on two other smaller DFT databases and an experimental database.

In this work, we introduce an inductive transfer learning approach with the transfer of model parameters from models trained on quantum chemistry calculations to models trained on experimental data, similar to the approach reported by Jha et al. \citep{Jha2019EnhancingLearning} The transfer learning algorithm uses inductive biases from the quantum chemistry data to improve learning of small and biased experimental datasets. In the present work, the transfer learning method is applied to the prediction of solvation free energies in a variety of solvents. For the purpose of this work we provide two databases in supporting information: (i) \revision{CombiSolv-QM -} a quantum chemistry database with solvation free energies for 1 million solvent/solute combinations, calculated according to the COSMO-RS theory, and (ii) \revision{CombiSolv-Exp -} an experimental database with 10145 solvent/solute combinations compiled from publicly available databases. For solvation free energies a rather large amount of data is available compared to other chemical or material properties. This allows us to investigate the effect of the dataset size and the influence of data noise or the aleatoric uncertainty on model predictions. 

Solvation free energies have been used before for the construction of deep neural networks. The FreeSolv database \citep{Mobley2014FreeSolv:Files}, with only hydration free energies, has often been used as a benchmark to compare different molecular representations in deep learning \citep{Wu2018MoleculeNet:Learning, Yang2019AnalyzingPrediction}. 
Some reported deep neural network architectures that account for multiple molecules are trained on data from the Minnesota Solvation database \citep{Marenich2012MinnesotaMNSOL}. Hutchinson and Kobayashi \citep{Hutchinson2019Solvent-SpecificLearning} used features for the representation of the solvent and functional class fingerprints for the solute. Lim and Jung \citep{Lim2019Delfos:Solvents} and Pathak et al. \citep{Pathak2020ChemicallyMolecules} proposed architectures with explicit solvent and solute embedding and an interaction layer to account for pair-wise interactions between solvent and solute latent representations. 
Both studies reported overall good predictions on random test splits with each an RMSE of 0.57 kcal/mol on their respective test sets. 
To test the transferability of the model to new solvents and solutes, Lim and Jung \citep{Lim2019Delfos:Solvents} clustered the solvents and solutes, re-trained the model while leaving out one cluster at a time, and tested performance on that cluster. On average, the RMSE of the model increased to 1.45 kcal/mol for solvent clustering and 1.61 kcal/mol for solute clustering. 
Pathak et al. \citep{Pathak2020ChemicallyMolecules} tested the transferability to new solvents in a similar manner by excluding certain solvents from the training set. They achieved overall good performance on test sets that include the left-out solvent. However, many of the considered solvents had chemical structures very similar to other solvents in the training set. 
In this work, the transferability of the model is tested by excluding specific solvents, solutes with certain elements, and solutes based on their molar mass from the training and validation sets. The transferability of the pre-trained models is compared to the transferability of the models trained solely on experimental data. 

\section{Methods - databases}

\subsection{\revision{CombiSolv-QM: the quantum chemical database}}

The Gibbs free energy of solvation ($\Delta G_{solv}$) at 298 K for generated data is determined for different solvent/solute combinations using the commercial software COSMO\textit{therm} \citep{DassaultSystemes2020BIOVIACOSMOtherm}. 
COSMO\textit{therm} computes thermophysical data of liquids based on the COSMO-RS theory \citep{Klamt1995Conductor-likePhenomena, Klamt1998RefinementCOSMO-RS, Eckert2002FastApproach}. Using this software, $\Delta G_{solv}$ is calculated from the chemical potential of the solute in the ideal gas phase and at infinite dilution in the considered solvent. The chemical potential is determined by considering pair-wise interactions between segments of the quantum chemical COSMO-surfaces of the solute and solvent molecules. 
The COSMO-surfaces used in this work are computed at BP-TZVPD-FINE level of theory, \textit{i.e.} using a geometry optimization on the density functional theory BP-TZVP level, a single point calculation on the BP-def2-TZVPD level and a FINE cavity for the construction of the surface segments. Different molecular conformations for the solvent and solute are accounted for in the calculation of $\Delta G_{solv}$. The conformer generation workflow considers conformers relevant for thermodynamic properties in the gas and in the liquid phase, as implemented in the commercial software COSMO\textit{conf}. The $\Delta G_{solv}$ values in this work are calculated in the molar reference state, meaning that $\Delta G_{solv}$ is the free energy for transferring a solute molecule from the ideal gas phase at 1 mol/L concentration into an ideal solution at the same solute concentration.

One of the advantages of using COSMO\textit{therm} for the calculation of $\Delta G_{solv}$ is that once the expensive quantum calculations are done for a new molecule and its conformers, the COSMO-surface of the new species can be quickly combined with already available COSMO-surfaces. This allows fast computation of new solvent/solute combinations to extend the quantum chemical database according to the user's needs. 
For the purpose of this work, a database is generated with 1 million combinations of 284 commonly used solvents and 11029 solutes. \revision{Those 1 million data points are randomly selected from all possible solvent-solute combinations.} Solvents and solutes with elements H, B, C, N, O, F, P, S, Cl, Br and I are included with a solute molar mass ranging from 2.02 g/mol to 1776.89 g/mol. All calculations are performed with resources of the National Energy Research Scientific Computing Center (NERSC). The complete \revision{CombiSolv-QM} database, with solvents and solutes represented by SMILES, can be found in the Supporting Information. \revision{CombiSolv-QM} contains 1 million solvation free energies for different solvent/solute combinations and can be used to further optimize machine learning architectures that consider multiple molecules and pair-wise interactions between those molecules. 

\subsection{\revision{CombiSolv-Exp: the} experimental database}

To construct the experimental database employed in this work, experimental data from different sources are combined. Those sources include the Minnesota Solvation database (MNSol) \citep{Marenich2012MinnesotaMNSOL}, the hydration free energy database  published by Mobley et al. \citep{Mobley2014FreeSolv:Files} (FreeSolv), the  database published by Moine et al. \citep{Moine2017EstimationSolutes} (CompSol) and a collection of data published by the Abraham research group at University College London \citep{Grubbs2010MathematicalSolvents}. 
For each of those databases, the reported molecular identifiers are converted to computer-readable identifiers, SMILES and InChI. Only solvation free energies at temperatures of 298K ($\pm$2K) are accounted for.
When the gas-liquid solvation equilibrium coefficient $K$ or $log(K)$ is reported instead of $\Delta G_{solv}$, the former is converted by the relation \revision{$K = \exp(-{\Delta G_{solv}}/{RT})$}.

The database is further curated by removing ionic liquids and ionic solutes. In addition, only molecules with the same elements as available in the quantum chemical database are allowed.
Even though it would be beneficial to train a machine learning model on a larger and more diverse experimental dataset, to demonstrate the advantage of transfer learning, the self-solvation experimental data are excluded from the final experimental database.  

The data from different sources are combined and duplicate entries are averaged or removed from the dataset if the standard deviation is larger than 0.20 kcal/mol. By applying this constraint, 181 entries are removed from the database. The selected standard deviation to remove entries from the database is based on reported experimental uncertainty of 0.20 kcal/mol for $\Delta G_{solv}$ of neutral components \citep{Kelly2005SM6:Clusters, Nicholls2008PredictingChemistry, Geballe2010TheOverview}. For some components however, the experimental uncertainty is larger and can go up to 1 kcal/mol \citep{Guthrie2009AOverview}. The final database consists of 10145 solvent/solute combinations for 291 solvents and 1368 solutes. The \revision{CombiSolv-Exp} database is provided as part of the Supporting Information, excluding the data from the proprietary MNSol database. The number of entries from each of the respective data sources are given in Table \ref{tab:tables_refs}.

\begin{table}[htbp]
    \caption{Number of database entries from each of the data sources}
    \label{tab:tables_refs}
    \centering
    \begin{tabular}{l|c}
        \hline
        \textbf{Data source} & \textbf{Number of entries}\\
        \hline        
        MNSol & 2275 \\
        FreeSolv & 560 \\
        CompSol & 3548 \\
        Abraham & 6091 \\
    \end{tabular}
\end{table}

\subsection{Comparison of the experimental and QM database}
One of the advantages of using a QM database over an experimental one is the lack of bias towards certain solvents. Over 11.8\% of the experimental data has water as a solvent and more than 10.0\% is in hydrocarbon solvents such as hexane, heptane, octane, decane and hexadecane. In the QM database, the data is more evenly distributed between solvents, and they each appear in about 0.35\% of the data.

\begin{figure}[t!]
\centering
\includegraphics[width=1.0\textwidth]{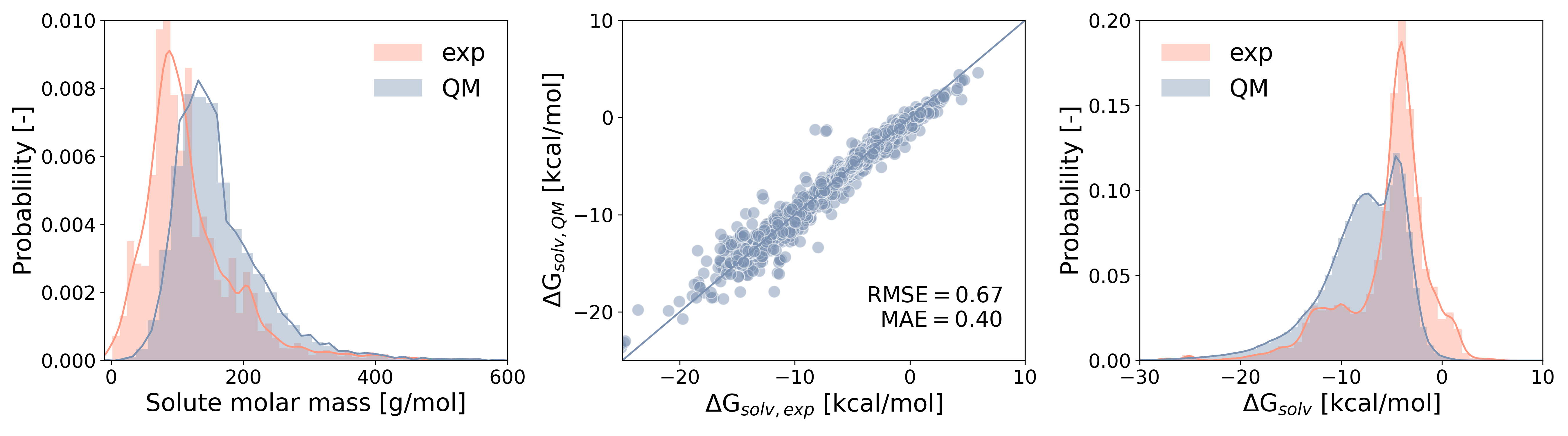}
\caption{Comparison of the experimental \revision{CombiSolv-Exp} (exp) and \revision{quantum chemistry CombiSolv-QM} (QM) database. left: molar mass distribution of the solutes, middle: parity plot for solvent/solute pairs which appear in both databases, right: distribution of $\Delta G_{solv}$.}
\label{fig:data}
\end{figure}

Another advantage of using a QM database is that more and new solutes can be considered compared to those that are experimentally available. The experimental database has 1368 different solutes, while 11029 different solutes are considered in the QM database. The molar mass distribution of solutes in the experimental database and the QM database are compared in Figure \ref{fig:data} (left). Note that the molar mass distribution is higher for the QM database. The component with the highest molar mass in the experimental database is hexabromobenzene with a molar mass of 551.49 g/mol, while in the QM database this is hexadecabromophthalocyanine with a molar mass of 1776.89 g/mol. The QM database can be extended at a computational cost but cheaper and faster than new experimental measurements to include more high molar mass components and to extend the application range of the model.

The values of the solvation free energies are compared between the experimental and QM database. There are 3164 overlapping solvent/solute combinations between the two databases. With the COSMO-RS theory, we can predict the experimental solvation free energy with a root-mean-square-error (RMSE) of 0.67 kcal/mol and a mean-absolute-error (MAE) of 0.40 kcal/mol.  This is close to the MAE reported by Klamt et al. \citep{Klamt2015CalculationDCOSMO-RS} of 0.42 kcal/mol for comparison to the SM8 dataset \citep{Cramer2008AModeling}. A parity plot can be seen in Figure \ref{fig:data} (middle).

The distribution of $\Delta G_{solv}$ for both databases is given in Figure \ref{fig:data} (right). The distribution is similar between the experimental and QM database, although a broader distribution is observed for the QM database. \revision{The maximum values found for $\Delta G_{solv}$ are 9.26 kcal/mol in the QM database and 5.92 kcal/mol in the experimental database. The lowest values for $\Delta G_{solv}$ are -58.13 kcal/mol and -47.92 kcal/mol in the QM and experimental database respectively. Note that the range values that are covered by $\Delta G_{solv}$ is smaller compared to, for example, the formation enthalpies of these molecules.}

\section{Methods - machine learning}
\subsection{Model architecture}
The model architecture used in this work is based on the state-of-the-art directed message passing neural network (D-MPNN) as implemented in the software chemprop. Only details specific to this work are discussed below. For more general information on the D-MPNN, the reader is referred to the work published by Yang et al. \citep{Yang2019AnalyzingPrediction} The software is written using the package PyTorch and is available open-source at https://github.com/chemprop/chemprop.

A molecular identifier, SMILES or InChI, of the solvent and solute molecule are converted into a graph-based structure by the open-source cheminformatics software RDKit \citep{2020RDKit:Cheminformatics}. 
Atom and bond feature vectors are constructed for each of the atoms and bonds in the graph representations. Those feature vectors are adapted from the standard version of chemprop to make them more specific to solvation related properties. The atom feature vectors contain information on (i) the atomic number, (ii) the number of neighboring atoms, (iii) the formal charge, (iv) the number of connected hydrogen atoms, (v) the hybridization, (vi) the number of lone pairs, (vii) the hydrogen bond donating or accepting character, (viii) the ring size, (ix) the aromaticity, (x) the electronegativity and (xi) the atomic molar mass. The bond feature vectors contain information on (i) the bond type, (ii) the conjugation, (iii) the ring type and (iv) the stereo-chemistry. The values of the atom and bond features are assigned by RDKit. 

\begin{figure}[h]
\centering
\includegraphics[width=0.6\textwidth]{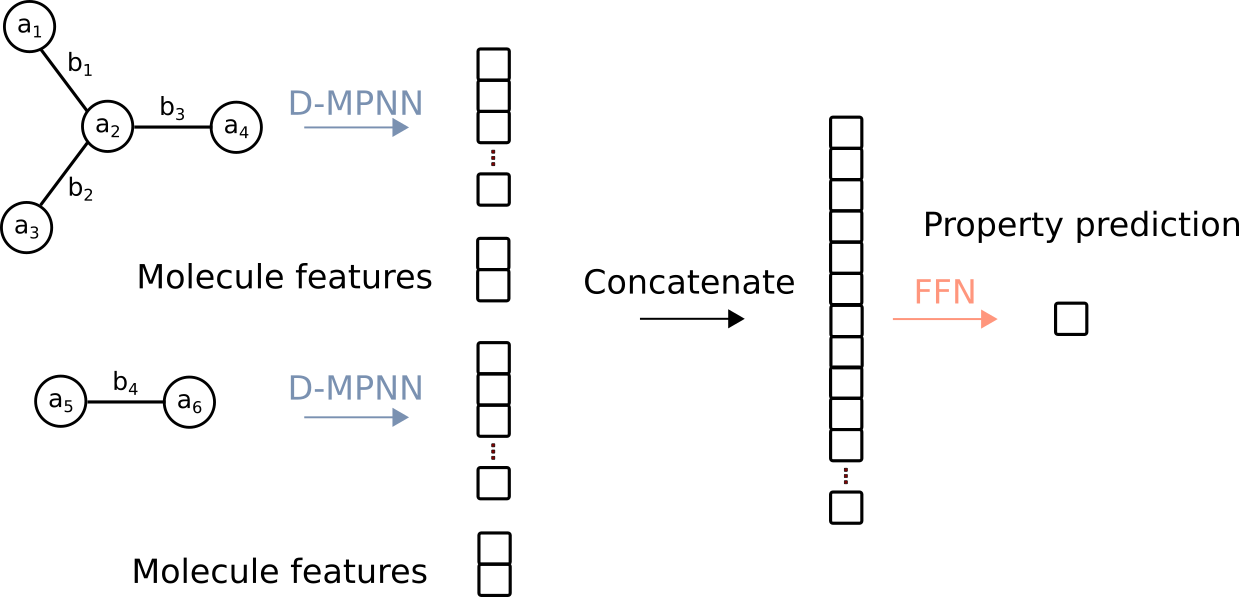}
\caption{Demonstration of the directed message passing neural network (D-MPNN) for solvent and solute molecular representation and the subsequent feed forward network (FFN) for property prediction}
\label{fig:mpn}
\end{figure}

The feature vectors are converted into a molecular latent representation by passing them through a convolutional neural network, more specifically a D-MPNN. This is done separately for the solvent and solute molecules and the latent representations of both molecules are concatenated. Some additional molecular features, such as the RDKit-calculated topological polar surface area and the RDKit-calculated molecular size, are concatenated with the molecular latent representations of the solvent and solute to improve the predictions. The concatenated embedding for the solvent and solute molecules are passed through a second neural network for the property prediction that is made up of linear feed forward layers. A schematic overview of the model architecture is given in Figure \ref{fig:mpn}.
Note that the D-MPNN and the network for property prediction, further referred to as the feed forward network (FFN), are treated differently during transfer learning.

Since this work focuses on demonstrating the benefits of transfer learning rather than optimizing the neural networks for property prediction, many of the hyperparameters for the model architecture and optimization of the neural network are fixed. \revision{Those hyper-parameters are selected manually based on previously performed hyperparameter optimizations with similar databases, while aiming at a small model architecture with a good model performance.}
For the D-MPNN, the depth of the message passing is set to 4 and the size of the hidden layers to 200 for both solvent and solute embedding. The D-MPNN linear layers have no bias and the best results are obtained without considering dropout. A LeakyReLU activation function is used to connect the different layers of the neural network. 
For the FFN, 4 layers are considered each with a hidden size of 500. The linear layers have a bias, no dropout, and are also connected with a LeakyReLU activation function.

Prior to training the neural network, all targets are normalized using the standard score. The parameters of the neural network are initialized randomly by a normal distribution as published by Glorot et al. \citep{Glorot2010UnderstandingNetworks}, except in the case of transfer learning where the model parameters are initialized using parameters from the other neural network.
Training of the neural network is done in batches of 50 datapoints for 200 epochs. A Noam learning rate scheduler is used with piece-wise linear increase and exponential decay, based on the learning rate scheduler in the Transformer model for Natural Language Processing \citep{NIPS2017_3f5ee243}. The model parameters are optimized with stochastic optimization as implemented in the Adam algorithm \citep{kingma2015adam} and based on the mean-squared-error loss.    
All models are trained on Nvidia Volta V100 GPUs on MIT SuperCloud \citep{reuther2018interactive}.

\subsection{Transfer learning}
\begin{figure}[tb]
\centering
\includegraphics[width=0.6\textwidth]{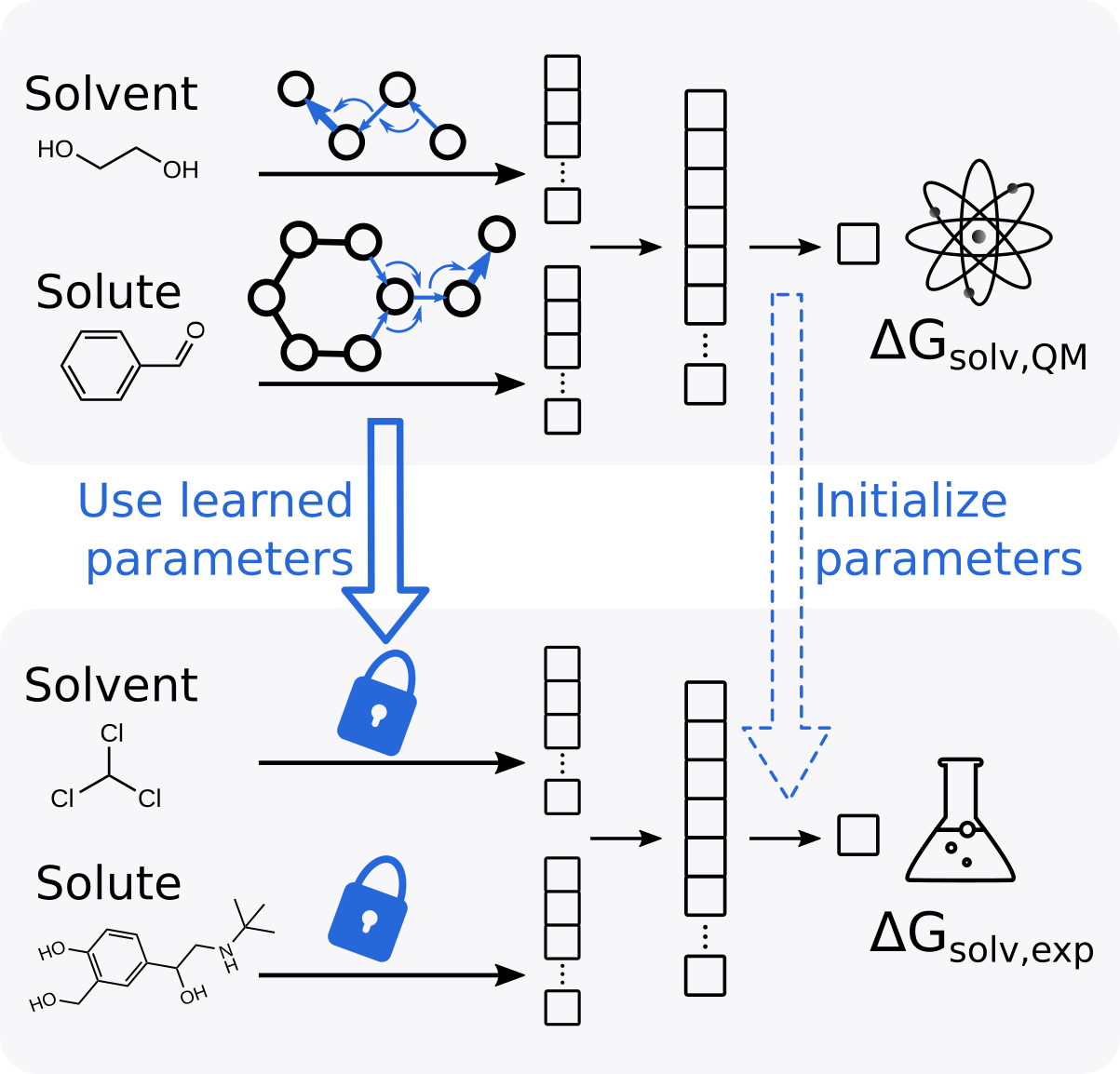}
\caption{Schematic representation of the transfer learning approach. The learned parameters from the directed message passing neural network (D-MPNN) are transferred from the model trained on the quantum chemical calculations to the new model fine-tuned on experimental data. The learned parameters of the feed forward network (FFN) for property prediction are used to initialize the parameters of the new model trained on experimental data.}
\label{fig:method}
\end{figure}

The design of the transfer learning approach is similar to the implementation by Grambow et al. \citep{Grambow2019AccurateApproach} In this work, the parameters of the models trained on \revision{the CombiSolv-QM} data (the QM models), are used to initialize the parameters of new models that are further refined using \revision{the CombiSolv-Exp} data (the pre-trained models). The parameters of the D-MPNN are frozen during optimization, while the parameters of the FFN are allowed to optimize for 20 epochs. The methodology is schematically presented in Figure \ref{fig:method}.
In later sections, the performance of these pre-trained models is compared to models trained on only experimental data (the experimental models). \revision{The pre-trained models and software are available through a conda package on https://anaconda.org/fhvermei/ml\_solvation.} 

\section{Performance of the quantum machine learning model}
The \revision{CombiSolv-QM} database is used to train 10 different models with 10-fold cross validation and random initiation of the model parameters. The train, validation, and test set are a 80/10/10 \% random split of the initial QM database. Further, the parameters from those 10 different models will be used for transfer learning. 
\subsection{Size of training data for the quantum chemistry model}
The final models for transfer learning are trained on the complete \revision{CombiSolv-QM} database with 1 million solvent/solute combinations. The rather large size of this new QM dataset provides an opportunity to investigate the influence of the size of the dataset to the model accuracy. To this purpose, the same training procedure is done starting from randomly selected subsets of the QM database.

\begin{figure}[]
\centering
\includegraphics[width=0.4\textwidth]{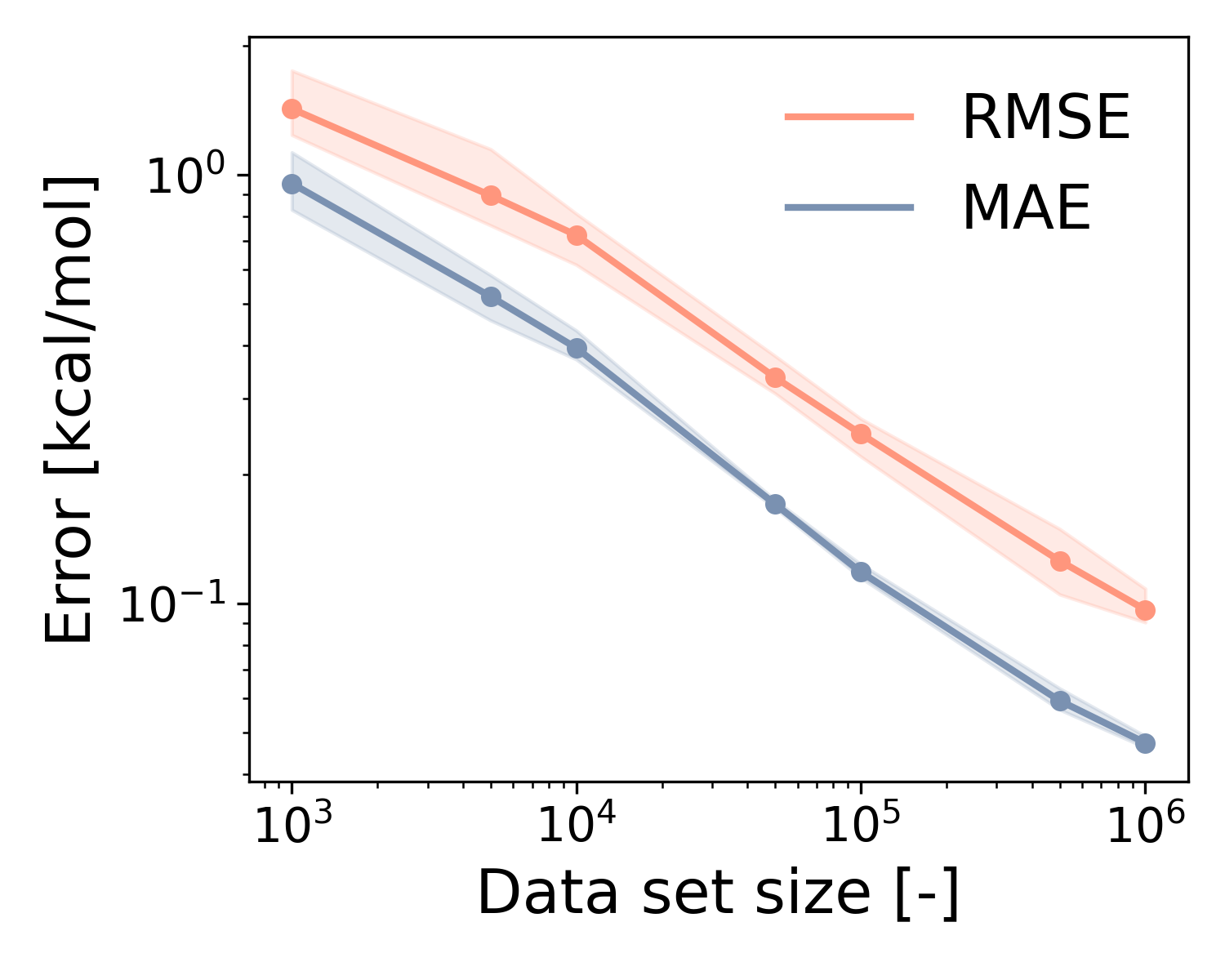}
\caption{Model for $\Delta G_{solv}$ trained and tested using varying amounts of COSMO-RS (quantum-derived) data. Root-mean-square-error (RMSE) and mean-absolute-error (MAE) in kcal/mol on a randomly selected 10\% test set as a function of the dataset size. The shaded area is defined by the maximum and minimum RMSE and MAE over the set of tested models.}
\label{fig:size_QM}
\end{figure}

The size of the dataset used for training, validation, and testing is varied between $10^{3}$ and $10^{6}$. 
The results of predictions on random test splits are presented in Figure \ref{fig:size_QM} as a function of the dataset size. The presented RMSE and MAE are the average of the RMSE and MAE of each of the models on their respective randomly selected test split. The uncertain shaded area in Figure \ref{fig:size_QM} is defined by the maximum and minimum RMSE and MAE over the set of tested models. 
The average RMSE/MAE are 1.43/0.95 \revision{kcal/mol} for $10^{3}$ data points, 0.72/0.39 \revision{kcal/mol} for $10^{4}$ data points, 0.25/0.12 \revision{kcal/mol} for $10^{5}$ data points and 0.10/0.05 \revision{kcal/mol} for $10^{6}$ data points. 
Similar trends for errors as a function of the dataset size were observed by von Lilienfeld et al. \citep{vonLilienfeld2020ExploringLearning} who reviewed predictions on the QM9 dataset and by Jha et al. \citep{Jha2018ElemNet:Composition} who performed deep learning for DFT enthalpies of formation for crystal structures. 
 
\subsection{Physical interpretation of solvents}
\begin{figure}[]
\centering
\includegraphics[width=0.9\textwidth]{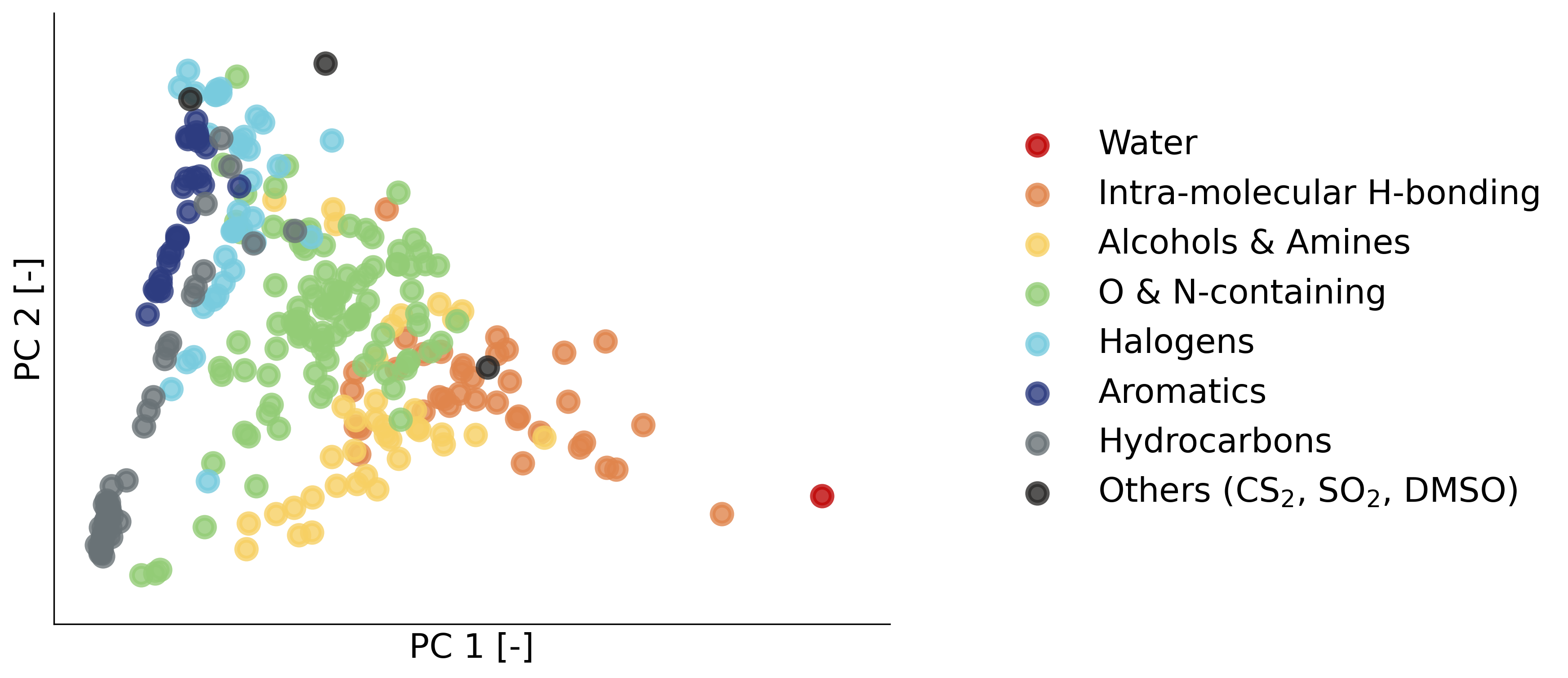}
\caption{First two principal components (PC1 and PC2) of the learned solvent molecular embedding. The solvents are classified manually based on the presence of features important to solvation in their chemical structure.}
\label{fig:pca_solvents}
\end{figure}

A principal component analysis (PCA) is performed on the solvent molecular embeddings for the 284 solvents in the \revision{CombiSolv-QM} database to demonstrate \revision{what is} learned by the QM model. \revision{Note that these results represent the learned behavior of a black-box model rather than exact physical properties, and the interpretation of those should be done with caution.} The first 2 principal components (PC1 and PC2) are graphically represented in Figure \ref{fig:pca_solvents}. Together they explain 41.9\% of the variance. 
The solvents are manually classified according to the presence of features specific to their chemical structure that are considered important for physical interactions in solvation, such as hydrogen bonding. Water is separated as a unique solvent, as well as carbon disulfide (CS$_2$), sulfur dioxide (SO$_2$), and dimethyl sulfoxide (DMSO). 
The first class, orange in Figure \ref{fig:pca_solvents}, includes all molecules that can have intra-molecular hydrogen bonding between oxygen and/or nitrogen atoms. The second class, yellow in Figure \ref{fig:pca_solvents}, includes solvents with one oxygen or nitrogen atom containing a hydrogen bond accepting and donating site. All other oxygen- and nitrogen-containing components are grouped in a third class, green in Figure \ref{fig:pca_solvents}.
The first three classes include some hetero-atomic aromatic structures, for example 2-furfuryl alcohol
, pyrolle
, phenol 
and pyridine
. Those types of solvents are, according to the first two principal components, better classified in the first three groups rather than based on their aromaticity. All other aromatics are included in a fourth group, indigo in Figure \ref{fig:pca_solvents}. The two remaining groups are the halogen-containing solvents, blue in Figure \ref{fig:pca_solvents}, and the hydrocarbon solvents, grey in Figure \ref{fig:pca_solvents}. 

PC1 distinguishes solvents based on their hydrogen bonding capability and the polarity of the solvents. The highest value of PC1 is for water
, followed by some molecules with the ability to make intra-molecular hydrogen bonding, such as propane-1,2-diol
, glycerol, 
and formamide
. 
The lowest values of PC1 are for a group of unsaturated hydrocarbon solvents in the lower left corner of Figure \ref{fig:pca_solvents}, including, for example, bicyclohexane 
and hexadecane
. 
PC2 separates molecules within one class based on the length of their hydrocarbon backbone. For example, within the halogens, carbon tetrachloride 
has the highest value of PC2, whilst the lowest values of PC2 for halogenes are for fluorooctane 
and bromooctane
. The alcohol and amine with the lowest value for PC2 are dodecanol 
and dibutyl amine 
respectively.

\subsection{Prediction of experimental data by the quantum chemistry models}
The QM models have an associated uncertainty when used to predict experimental solvation free energies. This absolute error is made up of the error inherent to the quantum chemical calculations and the additional error from the structure of the machine learning model. To evaluate those uncertainties, the 10 QM models are used to predict the complete experimental dataset without further fine-tuning of the model.

The predictions of the 10 models are averaged and the ensemble is used to calculate a RMSE of 0.81 kcal/mol and a MAE of 0.47 kcal/mol for the QM model predictions on experimental measurements. 
The absolute error of the predictions is expected to be close to the sum of the error of the QM calculations and the error of the model. The MAE of the former is 0.40 kcal/mol, see Figure \ref{fig:data} (middle), and the MAE of the latter is 0.05 kcal/mol as can be evaluated by the performance on a random test set, see Figure \ref{fig:size_QM}. 

\section{Transfer learning from quantum chemical calculations to experimental data}
The parameters of the QM models are used for transfer learning according to the procedure explained in Figure \ref{fig:method}. The model is fine-tuned on the complete set and subsets of the \revision{CombiSolv-Exp} dataset to demonstrate the advantage of transfer learning for small dataset sizes. Furthermore, we elaborate on the limited accuracy that can be reached with the employed experimental dataset caused by the noise inherent to the experimental nature of the data (\textit{i.e.} the aleatoric uncertainty). 
\subsection{Random splits of the complete experimental dataset}
The uncertainty associated with the quantum chemical calculations can be reduced by fine-tuning the model on experimental data. The performance has been evaluated using 10-fold cross validation. Fine-tuning the parameters of the neural network for property prediction improves the RMSE/MAE from 0.81/0.47 kcal/mol to 0.44/0.21 kcal/mol. 

\begin{figure}[]
\centering
\includegraphics[width=0.8\textwidth]{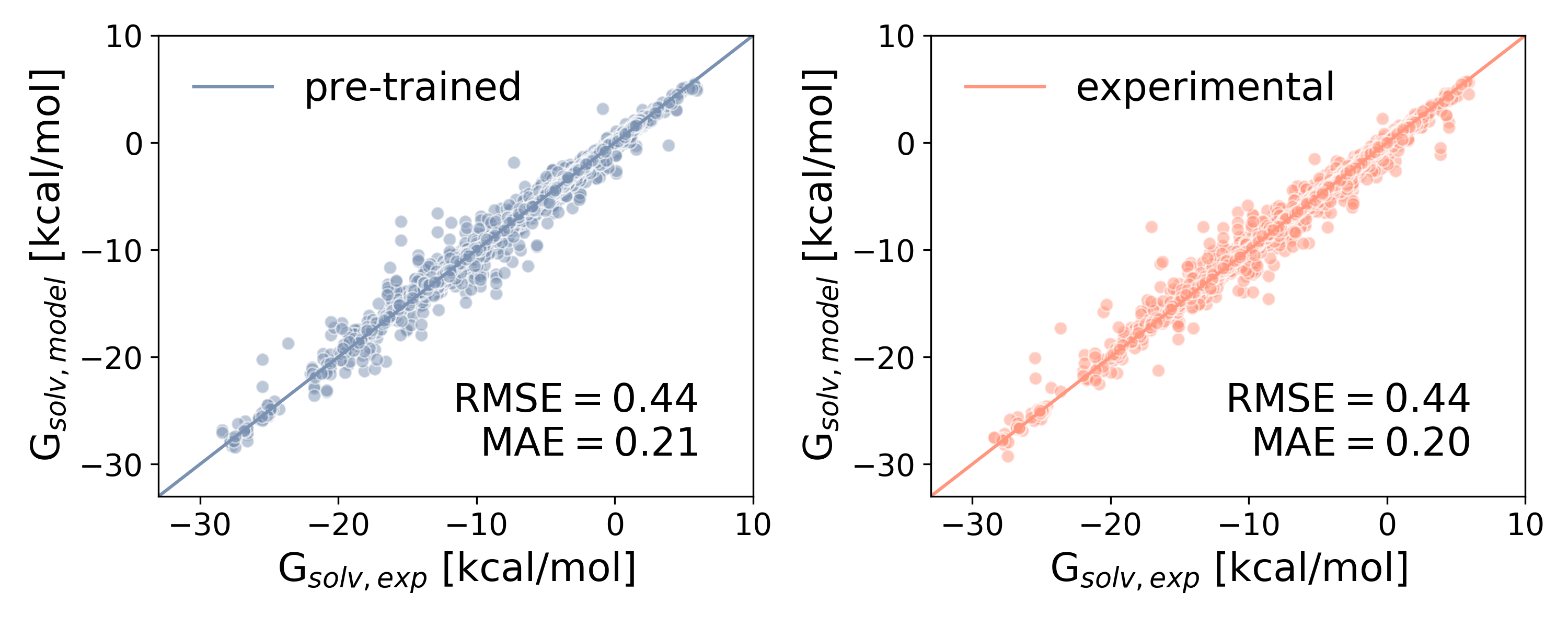}
\caption{Predictions of the pre-trained models (left) and the purely experimental models (right) on random splits of the experimental database.}
\label{fig:random_splits}
\end{figure}

For comparison with the pre-trained models, 10 experimental models are trained solely on the \revision{CombiSolv-Exp} database with random initiation of the model parameters. Those are trained using 10-fold cross validation to \revision{predict} the same random test splits as the pre-trained models. 
The predictions of the models on the random test splits can be seen in Figure \ref{fig:random_splits} for the pre-trained models (left) and the purely experimental models (right). Even though differences are observed between the predicted values, the overall performance of the experimental and pre-trained models are the same. 

\subsection{Influence of the experimental dataset size}
One of the advantages of transfer learning is the improved model performance on small datasets. For chemical and material properties, large datasets are often not available. For example, if a model would be trained for solvation free energies in specific solvents, for a common solvents such as dimethyl sulfoxide (DMSO) 
only 60 data points would be available. Moreover, in many industrial processes mixtures of solvents are used instead of one-component solvents. 
The more complex the solvent system gets, the more scarce the experimental data will be. Here we demonstrate the advantage of transfer learning with respect to small dataset sizes and that the same accuracy can be achieved as a model trained on solely experimental data for much smaller datasets. 

\begin{figure}[]
\centering
\includegraphics[width=0.8\textwidth]{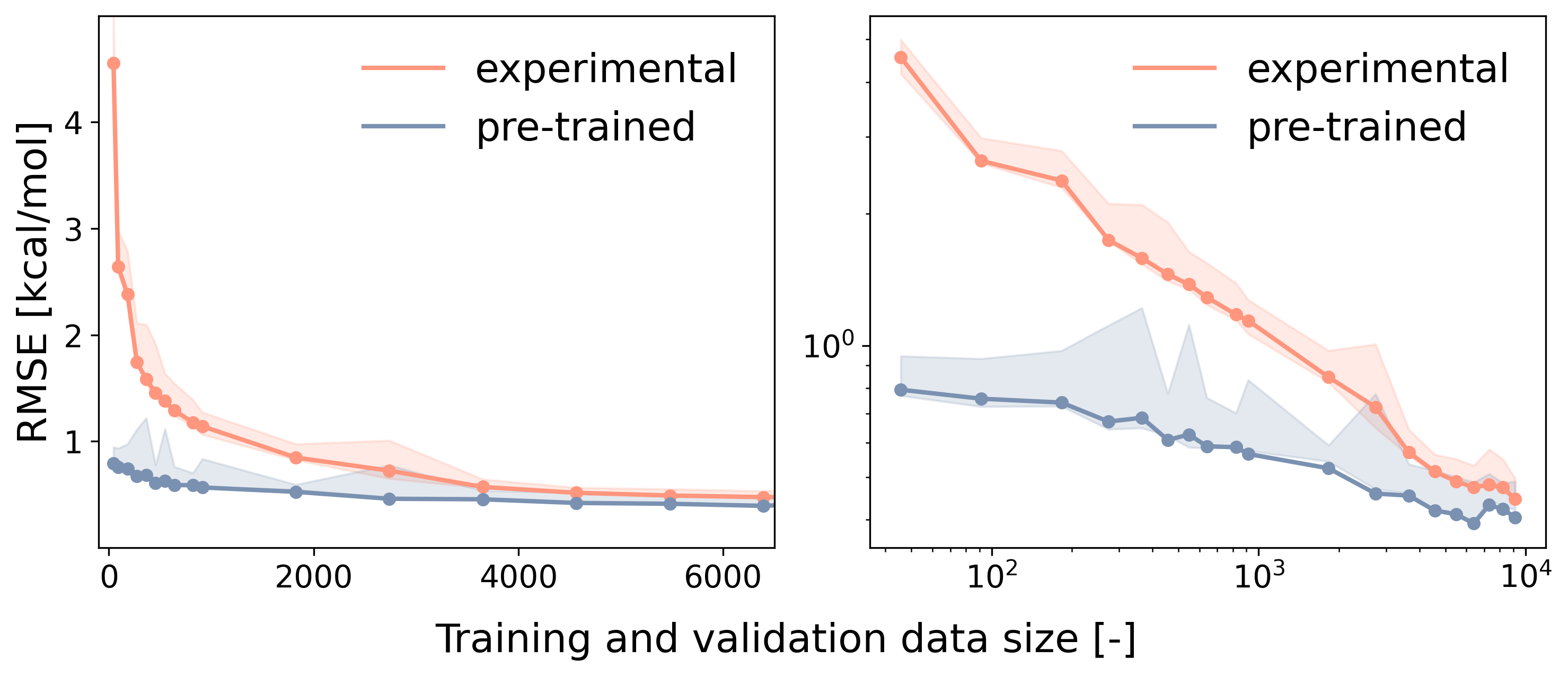}
\caption{Performance of the pre-trained (blue) and purely experimental (red) models on a fixed 10\% random test split as a function of the size of the training and validation set in linear (left) and logarithmic (right) scale. The shaded uncertainty area is defined by the minimum and maximum RMSE from each of the individual models in the ensemble.}
\label{fig:size_splits}
\end{figure}

A random 10\% of the experimental dataset is held out and used as separate test set. The size of the training and validation dataset is varied between 0.5\% and 70\% of the remaining experimental data, corresponding to 46 and 6392 data points respectively. 
In Figure \ref{fig:size_splits} the RMSE of the pre-trained and purely experimental models on the separate 10\% test set (1014 data points) are given as a function of the training and validation set size. The reported RMSE is calculated from the ensemble of the different models.
The shaded uncertainty area in Figure \ref{fig:size_splits} is defined by the minimum and maximum RMSE from each of the individual models in the ensemble.

Figure \ref{fig:size_splits} shows the decreasing error as a function of dataset size on a linear scale (left) and a logarithmic scale (right). The training data has significant noise leading to errors in the model predictions. This effect will be reduced by averaging when a large training dataset is used. A simplistic scaling argument says that effect should give a slope of -1/2 in Figure \ref{fig:size_splits} (right), in rough accord  with the observed slope for the purely experimental models.

The performance of the purely experimental models significantly increases with increasing size of the experimental dataset for the smallest fractions. This is the case until $\sim$50\% of the initial training set size, or $\sim$4567 data points. Once more data is added, the RMSE is slightly improved from 0.52 kcal/mol to 0.47 kcal/mol, but no significant further improvement of the model is observed. The pre-trained models perform significantly better for smaller dataset sizes. The same accuracy of 0.52 kcal/mol is reached at $\sim$20\% of the initial training set size, or $\sim$1826 data points. 
The resulting RMSE on a separate test set for 1\%, 5\% and 10\% of the initial training data (or 91, 457 and 913 data points) is improved from 2.64, 1.45 and 1.14 kcal/mol for the purely experimental models to 0.76, 0.61 and 0.57 kcal/mol for the pre-trained models.

\subsection{Aleatoric uncertainty as a limit to model performance}
When working with experimental datasets, the uncertainty related to the data noise should be accounted for in addition to the uncertainty associated with quantum chemical calculations and the model-related uncertainty (\textit{i.e.} epistemic uncertainty). 
Even a perfect model will have deviations from the test data due to noise in those data (\textit{i.e.} the aleatoric uncertainty). 
The aleatoric uncertainty is inherent to the experimental dataset and limits the accuracy of predictions on that data. This limit can be observed clearly in Figure \ref{fig:random_splits} and Figure \ref{fig:size_splits}. A lower limit of the model performance on the experimental test set with a MAE around 0.21 kcal/mol is approached by both the experimental and pre-trained models. This limit is the result of the aleatoric uncertainty in the dataset shared by both models. Note that the quantum chemistry dataset also has an aleatoric uncertainty, but much smaller than the one of the experimental dataset. 

As a demonstration of how noise inherent to the dataset can affect the prediction results, a subset of the experimental dataset with more accurate experimental measurements is constructed. 
Entries in the database that exhibit a low degree of variability are chosen. Those included have at least 3 unique measurements and a standard deviation below 0.15 kcal/mol.
This results in a new experimental dataset with 629 solvent/solute combinations for 82 solvents and 166 solutes. More than 20\% of the entries has water as a solvent and more than 10\% has linear hydrocarbon solvents. This data is used as a separate test set, while the QM models and the purely experimental models are retrained on the remaining experimental data (\textit{i.e.} excluding the new, more accurate test set). 
The RMSE and MAE on the test set are calculated based on the model ensembles. For a more accurate experimental test set, the RMSE/MAE can be reduced from 0.44/0.21 kcal/mol to 0.14/0.09 kcal/mol for the pre-trained models and to 0.18/0.10 kcal/mol for the purely experimental models.The aleatoric uncertainty or noise inherent to the experimental nature of the test set, limits the error that can be achieved for the predictions on this test set.
This result implies that the models trained on large amounts of data are significantly more accurate than one might infer from Figure \ref{fig:random_splits} and Figure \ref{fig:size_splits}. Most of the error seen in those figures is due to noise in the test set, not due to error in the model prediction. However, in the absence of less noisy test data it is impossible to quantify exactly how accurate the models are.

\section{Transfer learning to improve out-of-sample predictions}
On random test splits, the advantage of transfer learning is clear for small experimental dataset sizes. Besides this advantage, they are also expected to perform better on out-of-sample predictions. 
To demonstrate the advantage of a more physical model on a different region of chemical space, the QM models are fine-tuned on different pre-defined splits of the experimental data and compared to models trained on solely experimental data. The pre-trained models are constructed according to the transfer learning procedure used before (Figure \ref{fig:method}), and the experimental models are constructed with random initiation of the model parameters. The model parameters are refined with a pre-defined split of the experimental dataset and used to predict $\Delta G_{solv}$ of pre-defined test sets. The reported RMSEs and MAEs are based on the model ensembles. 

\subsection{Solvent splits}
To test the transfer learning methodology for out-of-sample solvents, some commonly used solvents are left out of the experimental training and validation dataset. The pre-trained and experimental models are trained with 10 different model initiations on the reduced experimental dataset and used to predict a test set that contains all experimental data of the left-out solvent. \revision{The previously trained QM models are used to make predictions for the same test set.}
Note that depending on the left-out solvent, the training, validation, and test set differ in size. The results are summarized in Table \ref{tab:table_solvents}. 

For some solvents, such as hexane
, acetone
, ethanol
, and benzene
, the performance of the pre-trained models and the experimental models are similar. Each of those solvents are part of a class of solvents that has a high occurrence in the experimental database. Many other hydrocarbon, ketone, alcohol, and aromatic hydrocarbon solvents are present in the experimental database. This demonstrates that the D-MPNN is effective in learning similarity in chemical structures even starting from an experimental database.
For other solvents like ethylacetate
, dichloromethane (DCM
), and acetonitrile
, the pre-trained models have an improved performance compared to the experimental models. These solvents also have counterparts in the experimental database with similar structures but not as many as the first set of solvents.
At last, for solvents with a more special chemical structure such as tetrahydrofuran (THF
), dimethyl sulfoxide (DMSO
), and water
, the pre-trained models outperform the experimental models significantly. 
\revision{For all left-out solvents, the pre-trained models outperform the QM models that are not further fine-tuned on experimental data, except for water. In the latter case, the performance of the QM models and pre-trained models is very similar.}

The comparison of predictions for out-of-sample solvents demonstrates that the D-MPNN is able to learn some similarities in chemical structures and make accurate predictions based on those similarities. However, for solvents with chemical structures less represented in the experimental training data, the performance of the D-MPNN diminishes. Transfer learning from QM data can significantly improve the model performance in this case.

When more accurate predictions are required for a new type of solvent, quantum calculations with this solvent can be added to the QM database at low computational cost. The pre-trained model can in this way be used to make predictions for the new solvent quickly and more accurately than either the direct application of the QM calculations or the experimental models.  

\begin{table}[htbp]
    \caption{Performance of \revision{QM, pre-trained, and experimental models} on out-of-sample solvent test sets. The reported RMSE and MAE are calculated on the predictions of the model ensembles. THF: tetrahydrofuran, DCM: dichloromethane, DMSO: dimethylsulfoxide}
    \label{tab:table_solvents}
    \centering
    \begin{tabular}{l|cc|cc|cc|c}
        \hline
        \multirow{2}{*}{\textbf{Solvent}} & 
        \multicolumn{2}{|c|}{\textbf{\revision{QM}}}&\multicolumn{2}{|c|}{\textbf{Pre-trained}} & \multicolumn{2}{|c|}{\textbf{Experimental}} & 
        \textbf{Test}\\
        
         & \textbf{\revision{RMSE}} & \textbf{\revision{MAE}} & \textbf{RMSE} & \textbf{MAE} & \textbf{RMSE} & \textbf{MAE} & \textbf{size} \\
        \hline
        hexane & \revision{0.37} & \revision{0.25} & 0.27 & 0.13 & \textbf{0.24} & \textbf{0.13} & 201 \\
        acetone & \revision{1.08} & \revision{0.61} & 0.24 & 0.15 & \textbf{0.23} & \textbf{0.18} & 100 \\
        ethanol & \revision{1.02} & \revision{0.67} & \textbf{0.50} & \textbf{0.20} & 0.51 & 0.27 & 144 \\
        benzene & \revision{0.82} & \revision{0.38} & 0.58 & 0.25 & \textbf{0.51} & \textbf{0.23} & 105 \\
        ethylacetate & \revision{0.92} & \revision{0.60} & \textbf{0.21} & \textbf{0.13} & 0.31 & 0.17 & 131 \\
        DCM & \revision{0.85} & \revision{0.44} & \textbf{0.25} & \textbf{0.16} & 0.34 & 0.21 & 51 \\
        acetonitrile & \revision{0.39} & \revision{0.24} & \textbf{0.13} & \textbf{0.10} & 0.23 & 0.18 & 67 \\
        THF & \revision{0.77} & \revision{0.48} & \textbf{0.24} & \textbf{0.17} & 0.47 & 0.37 & 116 \\
        DMSO & \revision{0.44} & \revision{0.37} & \textbf{0.34} & \textbf{0.27} & 0.95 & 0.85 & 60 \\
        water & \textbf{\revision{1.24}} & \textbf{\revision{0.77}} & 1.31 & 0.82 & 3.93 & 2.93 & 1200 \\
    \end{tabular}
\end{table}

\subsection{Solute splits}
The advantage of transfer learning on out-of-sample solutes is demonstrated using two different types of splits. The first split type is extreme, where certain elements are left out of the experimental training and validation set. 
The second split is a more commonly used split, where the models are trained on $\Delta G_{solv}$ for low molar mass solutes and used to predict those of high molar mass solutes. Extension of the chemical space to include solutes with a higher molar mass is useful for the determination of dry partitioning coefficients and solid solubility. Latter properties are valuable in the design of purification steps or for the selection of optimal solvents in, for example, drug discovery and synthesis. Moreover, gas-liquid solvation free energies for high molar mass components are often more difficult to measure experimentally and have a higher aleatoric uncertainty as a result. 

\subsubsection{Element-based splits}
In the out-of-sample solute element split, the model performance is evaluated on unknown elements. Solutes with those elements are left out of the experimental training and validation dataset and used to test the model performance. Depending on the identity of the excluded solute element, the training, validation, and test sets vary in size. The results are summarized in Table \ref{tab:table_solutes}. 
\begin{table}[htbp]
    \caption{Performance of \revision{QM, pre-trained, and experimental models} on out-of-range solute element test sets. The reported RMSE and MAE are calculated on the ensemble predictions.}
    \label{tab:table_solutes}
    \centering
    \begin{tabular}{l|cc|cc|cc|c}
        \hline
        \textbf{Excluded} & 
        \multicolumn{2}{|c|}{\textbf{\revision{QM}}}&\multicolumn{2}{|c|}{\textbf{Pre-trained}} & \multicolumn{2}{|c|}{\textbf{Experimental}} & 
        \textbf{Test}\\
        
        \textbf{element} & \textbf{\revision{RMSE}} & \textbf{\revision{MAE}} & \textbf{RMSE} & \textbf{MAE} & \textbf{RMSE} & \textbf{MAE} & \textbf{size} \\
        \hline
        
        O & \revision{0.96} & \revision{0.56} & \textbf{0.91} & \textbf{0.52} & 1.97 & 1.37 & 4684 \\
        N & \revision{1.13} & \revision{0.70} & \textbf{1.11} & \textbf{0.62} & 2.13 & 1.45 & 1559 \\
        F & \revision{0.99} & \revision{0.67} & \textbf{0.84} & \textbf{0.60} & 3.66 & 3.17 & 363 \\
        S & \revision{0.99} & \revision{0.68} & \textbf{0.94} & \textbf{0.64} & 1.71 & 1.34 & 369 \\
        Cl & \revision{0.75} & \revision{0.47} & \textbf{0.63} & \textbf{0.45} & 0.83 & 0.54 & 1124 \\
        Br & \revision{0.92} & \revision{0.55} & \textbf{0.51} & \textbf{0.28} & 0.56 & 0.41 & 216 \\
        I & \revision{0.60} & \revision{0.44} & \textbf{0.44} & \textbf{0.26} & 1.23 & 0.98 & 133 \\
    \end{tabular}
\end{table}

\revision{In all cases the pre-trained models outperform the experimental and QM models, since they learned the element representation in the neural network from the QM data and were further fine-tuned on experimental data.} Especially for elements that have hydrogen bonding accepting and donating sites, \textit{i.e.} O, N and F, the difference between the pre-trained and experimental models is significant. For more similar halogene elements, \textit{i.e.} Cl, Br and I, the gain from transfer learning is less significant. The experimental models are in this case able to learn some atom features, for example the electronegativity, from the similar halogene components. 

\subsubsection{Molar mass-based splits}
As could be seen in Figure \ref{fig:data} (left), the QM database includes more high molar mass solutes compared to the experimental database. Moreover, the QM database can be extended to include more high molar mass components, such that it covers more of the higher molar mass components for which $\Delta G_{solv}$ needs to be predicted. The advantage of transfer learning to predict $\Delta G_{solv}$ of out-of-sample, high molar mass, solutes is demonstrated by splitting the experimental database according to the solute molar mass. 

\begin{figure}[]
\centering
\includegraphics[width=0.4\textwidth]{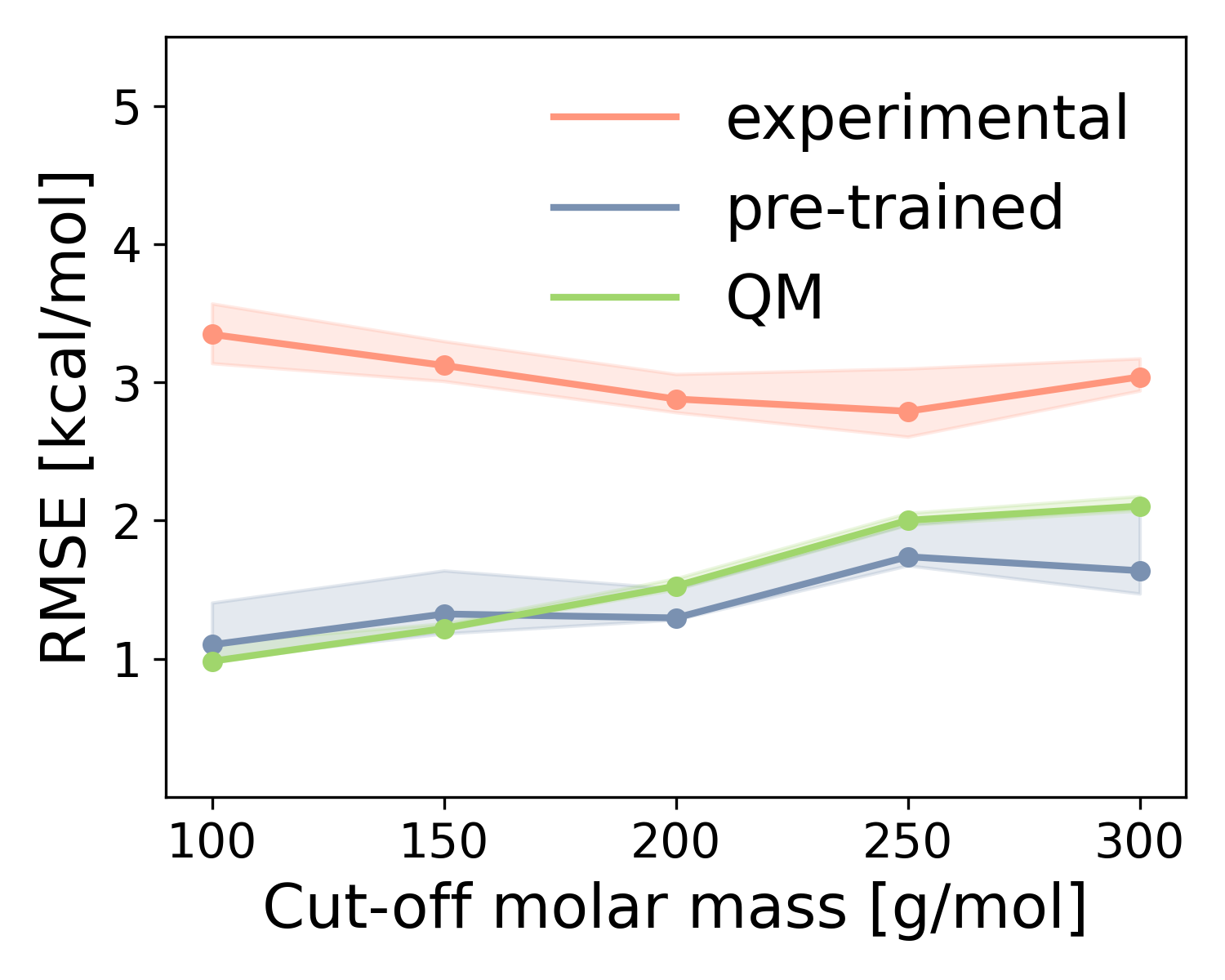}
\caption{Performance of the \revision{QM (green),} pre-trained (blue) and experimental (red) models on a test set with solutes that have a molar mass above the cut-off value. The \revision{pre-trained and experimental} models are trained on experimental data containing only solutes with a molar mass below the cut-off value. The shaded uncertainty area is defined by the maximum and minimum RMSE found for the individual models within the ensemble.}
\label{fig:MW_splits}
\end{figure}

The results for different molar mass splits are reported in Figure \ref{fig:MW_splits}. The training and validation dataset include all solutes with a molar mass below the cut-off value, while the model performance is tested for all solutes with molar mass above the cut-off value. The reported RMSE is determined from the model ensemble and the shaded uncertainty area is defined by the maximum and minimum RMSE found for the individual models within the ensemble. The size of the training and validation dataset differs between 4656 and 9860 for a cut-off molar mass equal to 100 g/mol and 300 g/mol respectively. 

For all cut-off solute molar mass splits, the pre-trained models outperform the models trained solely on experimental data. \revision{The pre-trained models have a similar performance to the QM models for a cut-off molar mass of 100 and 150 g/mol, while they outperform the QM models on the higher cut-off molar mass splits.} The trend in the RMSE as a function the cut-off molar mass is defined by a combination of the increasing size of the training dataset and the increasing complexity (or higher molar mass) of the test set.
Note that experimental data for most of the highest molar mass components are hydration free energies, \textit{i.e.} the considered solvent is water. Typically, the model performance is worse when predicting hydration free energies compared to other solvents (see also Table \ref{tab:table_solvents}). As a result, the RMSE reported in Figure \ref{fig:MW_splits} is significantly higher compared to the RMSE reported on random splits.

\section{Conclusions}
A transfer learning approach is presented that combines the advantages of quantum chemical and experimental databases in machine learning. Transfer learning from the quantum chemistry improves model predictions for molecules outside the range of the experimental training data, and for all molecules if the experimental training set is small. This is demonstrated for the prediction of solvation free energies in a variety of solvents. A QM database \revision{(CombiSolv-QM)} is reported with COSMO\textit{therm} calculations for 1 million solvent/solute combinations. An experimental database \revision{(CombiSolv-Exp)} is compiled from different data sources for 10145 solvent/solute combinations.

For the purpose of this work, the directed message passing neural network (D-MPNN) as developed by Yang et al\citep{Yang2019AnalyzingPrediction} is extended to include multiple molecules. During transfer learning, the deep neural network is first trained using the QM database. The model parameters of the QM models are used to initialize the parameters of new models which are refined using the experimental database, while the parameters of the D-MPNN for the solvent and the solute are frozen. 

To demonstrate the improved performance of the pre-trained models for small experimental dataset sizes, fractions of the experimental database are used for training and predicting a fixed 10\% test set. Especially for small experimental datasets, with up to 2000 data points, a clear advantage of transfer learning is demonstrated. For larger experimental datasets, the accuracy that can be achieved is limited by the noise in the experimental dataset. The model becomes sufficiently accurate such that the deviations between the model predictions and the test set are limited by the noise in the experimental test set. A significant improvement in performance, measured by deviations between model predictions and test set data, is gained when a subset of accurate experimental measurements with 629 datapoints is used for testing. 
The excellent performance of the model at predicting this high-accuracy data suggests that the model predictions of $\Delta G_{solv}$(298 K) have errors of less than $\sim$0.1 kcal/mol for in-scope molecules, but even more accurate test data would be needed to reliably determine this number.

The superior out-of-sample performance of the pre-trained models is demonstrated with pre-defined splits of the experimental data. For left-out solvents, left-out solute elements and a solute molar mass-based split, the pre-trained models outperform the purely experimental models. The experimental models are able to learn similarities in chemical structures compared to other entries in the database. However, for true out-of-sample data, the pre-trained models clearly have a superior performance.

\section*{Acknowledgements}
The authors acknowledge the Belgian American Educational Foundation (BAEF), the Machine Learning for Pharmaceutical Discovery and Synthesis Consortium (MLPDS), and the DARPA Accelerated Molecular Discovery (AMD) program (DARPA HR00111920025) for funding. This research used HPC resources to perform quantum calculation of the National Energy Research Scientific Computing Center (NERSC), a U.S. Department of Energy Office of Science User Facility operated under Contract No. DE-AC02-05CH11231.
The MIT SuperCloud Lincoln Laboratory Supercomputing Center is acknowledged for providing HPC resources to train the deep neural networks that have contributed to the research results reported within this paper. 
We thank Yunsie Chung for her help with the database construction, and Charles McGill and Lagnajit Pattanaik for useful discussions and revision of the manuscript.

\bibliography{references}

\end{document}